\documentclass[conference]{IEEEtran}
\ifCLASSINFOpdf
\else
\fi
\usepackage{graphicx}
\usepackage{setspace}
\usepackage{tablefootnote}
\usepackage{amsmath,empheq}
\usepackage{amssymb}
\usepackage{amsthm}
\usepackage{caption}
\usepackage{subcaption}
\usepackage{bm} 
\usepackage{cite}
\usepackage{float}
\usepackage[usenames,dvipsnames]{pstricks}
\usepackage{epsfig}
\usepackage{pst-grad} 
\usepackage{pst-plot} 
\usepackage{tabularx}
\usepackage[]{flushend}
\usepackage{amsmath}
\usepackage[T1]{fontenc}

\usepackage{xcolor}
\usepackage{lipsum}
\usepackage{comment}

\allowdisplaybreaks

\allowdisplaybreaks

\usepackage[left=1.62cm,right=1.62cm,top=1.9cm, bottom=1.9cm]{geometry}

\usepackage{caption}
\usepackage[utf8]{inputenc}

\begin{document}

\title{On Beamforming for Transmitter Location Privacy in MIMO Systems}

\author{
\IEEEauthorblockN{
Umair~Ali~Khan\IEEEauthorrefmark{1},
Lester~Ho\IEEEauthorrefmark{1},
Holger~Claussen\IEEEauthorrefmark{1}\IEEEauthorrefmark{2}\IEEEauthorrefmark{3},
Mark~F.~Flanagan\IEEEauthorrefmark{4},
and~Chinmoy~Kundu\IEEEauthorrefmark{1}
}

\IEEEauthorblockA{\IEEEauthorrefmark{1}Wireless Communications Laboratory, Tyndall National Institute, Dublin, Ireland}
\IEEEauthorblockA{\IEEEauthorrefmark{2}School of Computer Science and Information Technology, University College Cork, Cork, Ireland}
\IEEEauthorblockA{\IEEEauthorrefmark{3}Electronic \& Electrical Engineering, Trinity College Dublin, Dublin, Ireland}
\IEEEauthorblockA{\IEEEauthorrefmark{4}School of Electrical and Electronic Engineering, University College Dublin, Belfield, Ireland}

\IEEEauthorblockA{\{umairali.khan, lester.ho, holger.claussen, chinmoy.kundu\}@tyndall.ie, and mark.flanagan@ieee.org}
}

\maketitle
\pagestyle{plain} 
\begin{abstract}
In this paper, we introduce a beamforming framework to ensure transmitter location privacy against sensing-capable MIMO receivers. We propose a novel privacy metric called the \emph{direction-of-arrival obfuscation power ratio (DoA-OPR)} to design the transmit beamformer. The design approach reshapes the angular power distribution observed at the receiver so that a false direction appears dominant without nulling the line-of-sight (LoS) component. We derive closed-form bounds on the feasible range of DoA-OPR via generalized eigenvalue analysis and formulate an achievable rate-maximization problem under the DoA-OPR constraint. The resulting problem is non-convex but can be efficiently solved using semidefinite relaxation, eigenmode selection, and optimal power allocation.  Numerical results demonstrate that the proposed DoA-OPR-based beamformer achieves a trade-off between location privacy and communication rate. The proposed design attains higher achievable rates than existing LoS-nulling approaches while maintaining comparable location privacy. A suboptimal design strategy is also proposed with reduced complexity. It achieves a near-optimal communication rate with a reduction of nearly 85\% in computation time at a signal-to-noise ratio (SNR) of 10 dB.  
\end{abstract}

\vspace{-0.4cm}
\section{Introduction}

Applications such as industrial automation, smart building monitoring,
connected vehicles, and digital twin operation require
simultaneous high-throughput communication and accurate environmental
sensing. The integration of sensing capability in wireless networks can expose the transmitter's location, creating privacy concerns. 
Therefore, location privacy is vital to maintain trust in critical location-based services and ensure user safety  \cite{liu2024wifiSensing}.

Existing location privacy measures face various limitations; for example, upper-layer anonymization fails to thwart physical layer location inference attacks \cite{shokri2011quantifying}, and pilot-signal manipulation schemes are often incompatible with standardized pilot designs \cite{zhang2021robust}. While the transmit signal’s phase and amplitude change via repeaters and reconfigurable intelligent surfaces can impair an unauthorized receiver's direction-of-arrival (DoA) measurement ability, these techniques lack scalability due to infrastructure requirements and deployment costs \cite{xu2020privacy, taha2021irs}. Consequently, recent works in location privacy employ beamforming strategies to protect location information \cite{checa2022location,  ma2024sensingResistance,
ayyalasomayajula2023users,
zhang2023crb, 
li2023fakepath, 
tran2024fakepath}.

In \cite{checa2022location}, a mmWave transmit beamformer is designed to hide direction-of-departure (DoD) information of non-line-of-sight (NLoS) paths while suppressing the line-of-sight (LoS) path to the receiver, thus preventing the receiver from localizing the transmitter.  In \cite{ma2024sensingResistance}, a beamformer is designed to prevent sensing of the transmitter's real direction by suppressing the LoS component between the transmitter and the receiver.  
In \cite{ayyalasomayajula2023users},  
a transmit beamformer is designed to deceive a receiver into mistaking the NLoS path as the LoS path and thereby introducing localization error. This is achieved by adding a delay in the LoS path.
In \cite{zhang2023crb}, a multiple-input multiple-output (MIMO) orthogonal frequency division multiplexing based uplink localization system is considered where an unauthorized receiver intercepts a communication signal between a transmitter and a legitimate receiver and uses this to localize the transmitter. 
A transmit beamforming design is implemented to minimize the Cramér–Rao Bound (CRB) of legitimate localization while keeping the CRB of the unauthorized localization above a specified threshold. In \cite{li2023fakepath, tran2024fakepath}, fake-path injection employs transmit beamforming to synthesize virtual propagation paths that impair an illegitimate receiver’s DoA-based localization.

The articles  \cite{checa2022location,  ma2024sensingResistance,
ayyalasomayajula2023users}  consider a two-node scenario where a transmitter wants location privacy from a receiver with which it is communicating. However, articles \cite{zhang2023crb, 
li2023fakepath, 
tran2024fakepath} consider a three-node scenario where the transmitter wants location privacy from an illegitimate receiver, not from the receiver it is communicating with. In this paper, we consider the former case. In this case, while suppressing the LoS path between the transmitter and receiver, as in \cite{checa2022location,  ma2024sensingResistance},
can hide the true transmitter direction, the achievable rate is reduced. Although \cite{ayyalasomayajula2023users} does not suppress the LoS path, the technique used for location privacy can only be applied if the paths between the transmitter and receiver can be separated in the time domain and the direct path is identified. This may not always be possible. 

Motivated by the above discussion, we propose a novel location privacy metric called \emph{direction-of-arrival obfuscation power ratio} (DoA-OPR) to design the transmit beamformer that does not suppress the LoS and does not require identification of individual paths in the time domain. The proposed technique reshapes the angular power distribution such that the receiver observes more power in the false direction than in the true direction. 

The main contributions of this work are outlined as follows.
\begin{itemize}

\item A new privacy metric, termed DoA-OPR, is introduced, enabling a trade-off
between location privacy and communication rate without nulling the LoS path.  

    \item A transmit beamformer is designed to maximize the achievable rate under a location-privacy constraint (i.e., DoA-OPR), using joint eigenmode selection and power allocation.  
    \item A suboptimal solution (SS) strategy for the beamformer design is also proposed to reduce the computational complexity as compared to the optimal solution (OS) strategy with a slight performance loss.
    

    \item  Using the Capon DoA estimator, it is verified that the proposed design attains a privacy level equivalent to that of an existing design while attaining a higher rate.

\end{itemize}



\textit{Notation:}
$\mathbf{X}\sim\mathcal{CN}(\boldsymbol{\mu},\boldsymbol{\Sigma})$ denotes that the complex Gaussian random vector $\mathbf{X}$ has mean $\boldsymbol{\mu}$ and covariance matrix $\boldsymbol{\Sigma}$. $\mathbb{E}[\cdot]$ denotes expectation, $(\cdot)^{\mathrm{H}}$ denotes the Hermitian (conjugate) transpose, 
$\det(\cdot)$ denotes the determinant, $\operatorname{tr}(\cdot)$ denotes the trace, 
and $\mathbf{I}_N$ denotes the $N\times N$ identity matrix.

  \section{System and Channel Model}
\label{sec_system}
\begin{figure}[]
    \centering    \includegraphics[width=0.7\linewidth]{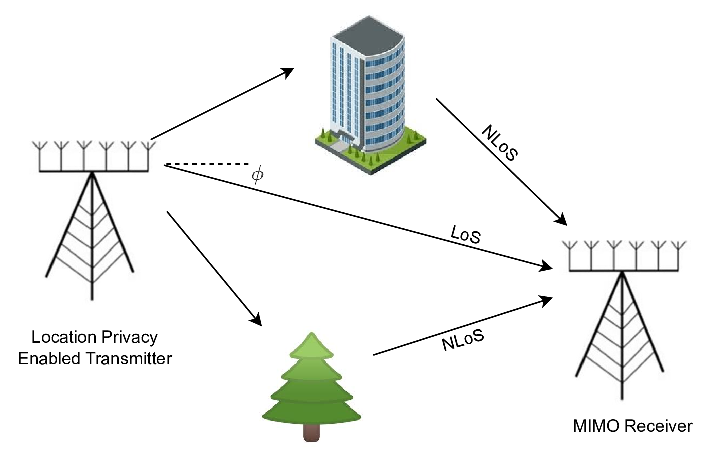} 
    \caption{Example of the location privacy enabled transmitter.}
    \label{fig_system_model}
\end{figure}

We consider a narrowband MIMO system where a transmitter equipped with $N_\textrm{T}$ antennas communicates with a receiver having $N_\textrm{R}$ antennas, as shown in Fig. \ref{fig_system_model}. We assume that both the transmitter and receiver employ antennas in the form of a uniform linear array (ULA). \footnote{The proposed privacy metric and beamforming formulation are applicable to other array architectures by substituting the appropriate array responses. } 
The normalized array steering vector for the transmit ULA can be written as \cite{JYuan_A_Tone_Based_AoA_Estimation,Robert_W_Heath_Channel_Estimation_and_Hybrid_Precoding}  
\begin{align}
\label{eq_steering_vector}
    \mathbf{a}_\textrm{T}(\phi) = \frac{1}{\sqrt{N_\textrm{T}}} \left[1, e^{-j\frac{2\pi d}{\lambda} \cos(\phi)}, \ldots, e^{-j\frac{2\pi d}{\lambda}(N_\textrm{T}-1)\cos(\phi)}\right]^T,
\end{align}
where $\phi\in [0,180]^\circ$ is the DoD, $d$ is the spacing between antennas, and $\lambda$ denotes the wavelength of operation. The array steering vector of the receiver, $\mathbf{a}_\textrm{R}(\phi)$, is defined analogously.


The transmitter employs a precoder to transform the power normalized complex baseband signal $\mathbf{s} \sim  \mathcal{CN}(\mathbf{0}, \mathbf{I}_{N_\textrm{S}}) \in  \mathbb{C}^{N_\textrm{S}}$ with $N_\textrm{S} \leq \min\{N_\textrm{T}, N_\textrm{R}\}$ data streams into the complex transmit signal $\mathbf{x}=\mathbf{W} \mathbf{s} \in \mathbb{C}^{N_\textrm{T}}$ where $\mathbf{W} \in \mathbb{C}^{N_\textrm{T} \times N_\textrm{S}}$ is the precoding matrix satisfying the transmit power constraint $\text{tr}(\mathbf{W}\mathbf{W}^H) = P$. The complex baseband received signal $\mathbf{y}\in  \mathbb{C}^{N_\textrm{R}}$ is then written as
\begin{align}
    \mathbf{y} = \mathbf{H} \mathbf{x} + \mathbf{n},
\end{align}
where  
$\mathbf{H} \in \mathbb{C}^{N_\textrm{R} \times N_\textrm{T}}$ represents the $N_\textrm{R} \times N_\textrm{T} $ channel matrix between the $N_\textrm{T} $ transmitting and $N_\textrm{R}$ receiving antennas and $\mathbf{n} \sim \mathcal{CN}(\mathbf{0}, N_0\mathbf{I}_{N_\textrm{R}})  \in \mathbb{C}^{N_\textrm{R}}$ is the complex additive white Gaussian noise (AWGN) vector. Here, $N_0$ is the variance of the AWGN at each receiving antenna.

The wireless channel is assumed to be a mmWave channel that follows the Rician fading model.  Incorporating both deterministic LoS and multipath NLoS components, it is expressed as \cite{Robert_W_Heath_Channel_Estimation_and_Hybrid_Precoding,JYuan_A_Tone_Based_AoA_Estimation}
\begin{align}
\label{eq_channel}
    \mathbf{H} = \sqrt{\frac{\kappa}{\kappa+1}} \mathbf{H}_{\text{LoS}} + \sqrt{\frac{1}{\kappa+1}} \mathbf{H}_{\text{NLoS}},
\end{align}
where $\mathbf{H}_{\text{LoS}}$ models the deterministic LoS component, and $\mathbf{H}_{\text{NLoS}}$ accounts for $L$ multipath components, and $\kappa$ is the  Rician K-factor, which is the ratio between the power of the LoS component and the power of the NLoS components. The normalized LoS component can be expressed using the relevant steering vectors as
\begin{align}
    \mathbf{H}_{\text{LoS}} = \sqrt{N_\textrm{T} N_\textrm{R}} \, \mathbf{a}_\textrm{R}(\phi) \mathbf{a}_\textrm{T}^H(\phi).
\end{align}
The NLoS components are modelled as
\begin{align}
    \mathbf{H}_{\text{NLoS}} = \sqrt{\frac{N_\textrm{T} N_\textrm{R}}{L}} \sum_{\ell=1}^{L} \alpha_\ell \, \mathbf{a}_\textrm{R}(\omega_{\textrm{r},\ell}) \mathbf{a}_\textrm{T}^H(\omega_{\textrm{t},\ell}),
\end{align}
where $\alpha_\ell \sim \mathcal{CN}(0,1)$ represents the complex small-scale fading coefficient for the $\ell$-th path normalized to unit variance, and $\omega_{t,\ell}, \omega_{r,\ell} \in [0,180]^\circ$ denote the uniformly distributed DoD and DoA of the $\ell$-th path. The scaling $\sqrt{N_\textrm{T} N_\textrm{R}}$ ensures power normalization per path.  
The channel model assumed in (\ref{eq_channel}) enables the capture of both directional LoS propagation and isotropic multipath scattering, making it suitable for evaluating beamforming strategies in realistic MIMO environments.
It is assumed that $\mathbf{H}$ is known at the transmitter. \footnote{A Rician channel model is adopted for illustration; however, the proposed beamforming framework can be applied to any MIMO channel model.}

The achievable rate $C$ for the beamformer $\mathbf{W}$ can be expressed as
\begin{align}
\label{eq_rate}
C = \log_2\det\left(\mathbf{I}_{N_\textrm{R}} + \frac{1}{N_0}\mathbf{H}\mathbf{W}\mathbf{W}^H\mathbf{H}^H\right).
\end{align}

\section{Location Privacy Enabled Beamforming}
\label{sec_beamforming}
In this section, we design the transmit beamformer to alter the angular energy distribution at the receiver, thereby misleading it regarding the transmitter’s actual direction while maximizing the achievable rate. The beamforming strategy exploits both LoS and NLoS paths to design a received beampattern that makes the false direction $\hat{\phi}$ appear more dominant than the true direction $\phi$, thereby achieving location privacy.
To achieve this, a new metric DoA-OPR is introduced, which we define as the ratio of the power in the false direction $\hat{\phi}$ to the power in the true direction $\phi$, which can be written as \begin{align}
\label{eq_DOAR}
\gamma = \frac{\mathbb{E}[(\mathbf{a}_\textrm{R}^H(\hat{\phi})\mathbf{y})(\mathbf{a}_\textrm{R}^H(\hat{\phi})\mathbf{y})^H]}
{\mathbb{E}[(\mathbf{a}_\textrm{R}^H(\phi)\mathbf{y})(\mathbf{a}_\textrm{R}^H(\phi)\mathbf{y})^H]} = \frac{\mathbf{a}_\textrm{R}^H(\hat{\phi})\mathbf{R}\mathbf{a}_\textrm{R}(\hat{\phi})}
{\mathbf{a}_\textrm{R}^H(\phi)\mathbf{R}\mathbf{a}_\textrm{R}(\phi)},
\end{align}
where 
$\mathbf{R}$ is the spatial covariance matrix of the received signal which is given by
\begin{align}
\mathbf{R} = \mathbb{E}[\mathbf{y}\mathbf{y}^H] =
\mathbf{H}\mathbf{W}\mathbf{W}^H\mathbf{H}^H +  N_0\mathbf{I}_{N_\textrm{R}}.
\end{align}
A value of $\gamma$ larger than 1 implies that the receiver is more likely to perceive $\hat{\phi}$ as the dominant incoming signal direction, thereby providing location privacy to the transmitter.



\subsection{Problem Formulation}
\label{sec_Problem_Formulation}
To design the beamformer, an optimization problem is  formulated which maximizes the achievable rate subject to the power constraint and a desired angular obfuscation level, i.e.,   
\begin{align}
\label{eq:opt_main}
\textrm{P}1:\quad  \max_{\mathbf{W}} \quad & C \\
\label{eq_power_constraint}
\text{s.t.} \quad &
\text{tr}(\mathbf{W}\mathbf{W}^H) = P,   \\
\label{eq_constraint_gamma}
& \gamma \geq \gamma_{\textrm{th}},  
\end{align}
where $\gamma_{\textrm{th}}$ is a pre-defined DoA-OPR (i.e., privacy) threshold. The first constraint enforces the total power limit, while the second ensures that the beamformer achieves a desired level of location privacy.

To investigate the feasibility of the optimization problem depending on $\gamma_{\textrm{th}}$ in P1, we first find the maximum and minimum values of $\gamma$, respectively. 
Towards this goal, (\ref{eq_DOAR}) can be rewritten as 
\begin{align}
\gamma &= \frac{\text{tr}(\mathbf{a}_\textrm{R}^H(\hat{\phi})\mathbf{R}\mathbf{a}_\textrm{R}(\hat{\phi}))}{\text{tr}(\mathbf{a}_\textrm{R}^H({\phi})\mathbf{R}\mathbf{a}_\textrm{R}({\phi}))} 
\label{er_gamma_rayleighQ_form}
=\frac{\text{tr}(\mathbf{W}^H \mathbf{A}_{\text{false}} \mathbf{W})}{\text{tr}(\mathbf{W}^H \mathbf{A}_{\text{true}} \mathbf{W})},
\end{align}
where 
\begin{align}
\mathbf{A}_{\text{false}} &= \mathbf{H}^H \mathbf{a}_\textrm{R}(\hat{\phi}) \mathbf{a}_\textrm{R}^H(\hat{\phi}) \mathbf{H} + \frac{N_\textrm{R} N_0}{P} \mathbf{I}_{N_\textrm{T}}, \\
\mathbf{A}_{\text{true}} &= \mathbf{H}^H \mathbf{a}_\textrm{R}(\phi) \mathbf{a}_\textrm{R}^H(\phi) \mathbf{H} + \frac{N_\textrm{R} N_0}{P} \mathbf{I}_{N_\textrm{T}}.
\end{align}
As $\mathbf{A}_{\text{false}}\in \mathbb{C}^{N_\textrm{T} \times N_\textrm{T}}$ and $\mathbf{A}_{\text{true}}\in \mathbb{C}^{N_\textrm{T} \times N_\textrm{T}}$ are positive definite and Hermitian, $\gamma$ in (\ref{er_gamma_rayleighQ_form}) is in the form of a generalized Rayleigh quotient. Thus, the feasible range of $\gamma$ is defined as
\begin{align}
\label{eq_limits_on_gamma}
\gamma_{\min} \leq \gamma \leq \gamma_{\max},
\end{align}
where $\gamma_{\min}$  and $\gamma_{\max}$ denote the smallest and largest generalized eigenvalue of the matrix pair \{$\mathbf{A}_{\text{false}}$, $\mathbf{A}_{\text{true}}$\}, respectively \cite{Channel_Estimation}.   The first and second equalities in (\ref{eq_limits_on_gamma}) hold when  $\mathbf{W}$ is designed using the generalized eigenvectors  corresponding to 
$\gamma_{\min}$  and $\gamma_{\max}$, respectively \cite{ma2024sensingResistance}. Specifically, 
the first and second equalities hold when 
\begin{align}
\label{eq_threshold_lowerbound}
\mathbf{W}\mathbf{W}^H = P\frac{\mathbf{t}_{\min}\mathbf{t}_{\min}^H}{\mathbf{t}_{\min}^H\mathbf{t}_{\min}},\\
\label{eq_threshold_upperbound}
\mathbf{W}\mathbf{W}^H  = P\frac{\mathbf{t}_{\max}\mathbf{t}_{\max}^H}{\mathbf{t}_{\max}^H\mathbf{t}_{\max}},
\end{align}
respectively, where 
$\mathbf{t}_{\min}$ and $\mathbf{t}_{\max}$ are the generalized eigenvectors corresponding to $ \gamma_{\min}$  and $\gamma_{\max}$, respectively.

Depending on whether the value of $\gamma_\textrm{th}$ lies within the range shown in (\ref{eq_limits_on_gamma}), the following cases arise for the solution of P1. 
\begin{enumerate}
    \item When $\gamma_\textrm{th}>\gamma_{\max}$: The solution of P1 is not feasible. In this case, the desired privacy constraint cannot be met by the system.
    \item When $\gamma_\textrm{th}=\gamma_{\min}$ or $\gamma_\textrm{th}=\gamma_{\max}$: The solution of P1 is obtained in closed form using  (\ref{eq_threshold_lowerbound}) or (\ref{eq_threshold_upperbound}), respectively.
     \item When $\gamma_\textrm{th}<\gamma_{\min}$: The privacy constraint in (\ref{eq_constraint_gamma}) is inactive so the conventional water-filling algorithm is used to obtain $\mathbf{W}$.

     \item When $\gamma_{\min} < \gamma_\textrm{th}<\gamma_{\max}$:  The problem P1 is non-convex due to the privacy constraint in (\ref{eq_constraint_gamma}); thus, it is difficult to find its solution directly. This is the main case of interest in this paper. The solution strategy of this case is described in the following subsection.
\end{enumerate} 



\subsection{Solution Strategy when $\gamma_{\min} < \gamma_{\normalfont \text{th}}<\gamma_{\max}$.}
In this case, the solution to the problem can be handled using the semidefinite relaxation (SDR) approach, assuming $\mathbf{Z}=\mathbf{W}\mathbf{W}^H \in  \mathbb{C}^{N_\textrm{T}\times N_\textrm{T}}$. The problem P1 can thus be  equivalently written as
\begin{align}
\label{eq_new_problem_P2}
\textrm{P}2:\quad \max_{\mathbf{Z}} \quad & \log_2 \det\left( \mathbf{I}_{N_\textrm{R}} + \frac{1}{N_0} \mathbf{H} \mathbf{Z} \mathbf{H}^H \right) \\
\label{eq_new_power_constraint}
\text{s.t.} \quad & \text{tr}
(\mathbf{Z}) = P, \\
\label{eq_new_provacy_constraint}
& \text{tr}\left( (\mathbf{A}_{\text{false}} - \gamma_\textrm{th} \mathbf{A}_{\text{true}})\, \mathbf{Z} \right) \geq 0, \\
\label{eq_rankZ}
& \mathbf{Z}=\mathbf{Z}^H \succeq \mathbf{0},\quad \text{rank}(\mathbf{Z}) = N_\textrm{S}.
\end{align}
In P2, (\ref{eq_new_power_constraint}) represents the total power constraint, (\ref{eq_new_provacy_constraint}) represents the privacy constraint and is obtained from  (\ref{eq_constraint_gamma}) using $\gamma$ in (\ref{er_gamma_rayleighQ_form}), and (\ref{eq_rankZ}) represents the constraint on $\mathbf{Z}$ which is a Hermitian  positive semidefinite matrix with rank $N_\textrm{S}$ due to its definition.

Note that $(\mathbf{A}_{\text{false}} - \gamma_\textrm{th} \mathbf{A}_{\text{true}})\in   \mathbb{C}^{N_\textrm{T} \times N_\textrm{T}}$ in (\ref{eq_new_provacy_constraint}) is Hermitian as it is a linear combination of Hermitian matrices.
Thus, 
performing eigenvalue decomposition on $(\mathbf{A}_{\text{false}} - \gamma_{\textrm{th}} \mathbf{A}_{\text{true}})$ we obtain
\begin{align}
\label{eq_eigen_modes}
\mathbf{A}_{\text{false}} - \gamma_{\textrm{th}} \mathbf{A}_{\text{true}} = \mathbf{U} \boldsymbol{\Lambda}  \mathbf{U}^H,
\end{align}
where $\mathbf{U}$ is unitary and $\boldsymbol{\Lambda}  = \text{diag}(\lambda_1,\lambda_2, \dots, \lambda_{N_\textrm{T}})$ is a diagonal matrix containing on its main diagonal the $N_\textrm{T}$ real eigenvalues $\lambda_1, \lambda_2, \ldots, \lambda_{N_\textrm{T}}$ of  $(\mathbf{A}_{\text{false}} - \gamma_{\textrm{th}} \mathbf{A}_{\text{true}})$. Looking at the product $(\mathbf{A}_{\text{false}} - \gamma_\textrm{th} \mathbf{A}_{\text{true}})\mathbf{Z}$ in (\ref{eq_new_provacy_constraint}) and considering $\mathbf{Z}$ is also a Hermitian matrix, we design $\mathbf{Z}$ in such a way that matrices $(\mathbf{A}_{\text{false}} - \gamma_\textrm{th} \mathbf{A}_{\text{true}})$ and $\mathbf{Z}$ commute to simplify the problem P2. In this case, matrices $(\mathbf{A}_{\text{false}} - \gamma_\textrm{th} \mathbf{A}_{\text{true}})$ and $\mathbf{Z}$ can be simultaneously diagonalized and share the same unitary matrix $\mathbf{U}$. Thus, we design
\begin{align}
\label{eq_SDR_Z}
\mathbf{Z} = \mathbf{U} \mathbf{P} \mathbf{U}^H,
\end{align}
such that $\mathbf{P} = \text{diag}(p_1, p_2, \ldots, p_{N_\textrm{T}})$ is a diagonal matrix with real non-negative entries, $\text{tr}
(\mathbf{P})  =  P$, and $\textrm{rank} (\mathbf{P}) = N_\textrm{S}$. To enforce the rank constraint on $\mathbf{P}$, out of $N_\textrm{T}$ diagonal entries of $\mathbf{P}$, only $N_\textrm{S}$ will be non-zero.

Taking into account (\ref{eq_eigen_modes}) and (\ref{eq_SDR_Z}), we simplify P2 as
\begin{align}
\label{eq_new_capacity}
\textrm{P3}:\max_{\small{\begin{array}{l}
            \mathcal{I}, \{ p_i \}_{i \in \mathcal{I}}
        \end{array}}}  & \log_2 \det \left( \mathbf{I}_{N_\textrm{R}} + \frac{1}{N_0} \mathbf{H} \mathbf{U}_{\mathcal{I}} \mathbf{P}_{\mathcal{I}} \mathbf{U}_{\mathcal{I}}^H \mathbf{H}^H \right) \\
\text{s.t.} \quad & \sum p_{i} = P, \\ 
\label{eq_constraint_product_privacy}
\quad &  \sum p_{i} \lambda_i\ge0,\\
\label{eq_subset}
  &~~\mathcal{I} \subseteq \{1, 2,\ldots, N_\textrm{T}\},\\ 
\label{eq_cardinality}
~~ &~~|\mathcal{I}|=N_\textrm{S},\\
~~ &~~p_i > 0 ~~\forall i \in \mathcal{I},
\end{align}
where $\mathcal{I}$ denotes the set (of size $N_\textrm{S}$) of indices  of non-negative entries of $\mathbf{P}$, 
and  $\mathbf{U}_{\mathcal{I}}$ and $\mathbf{P}_{\mathcal{I}}$ are the submatrices taking columns with indices in $\mathcal{I}$ from $\mathbf{U}$ and $\mathbf{P}$, respectively. The elements of $\mathbf{P}_{\mathcal{I}}$ are the allocated powers to the $N_\textrm{S}$ eigenvectors in (\ref{eq_eigen_modes}).
The constraint in (\ref{eq_constraint_product_privacy}) is obtained from (\ref{eq_new_provacy_constraint}) with the help of (\ref{eq_eigen_modes}) and (\ref{eq_SDR_Z}). We propose two solution strategies for P3: the OS and SS strategies.

\subsection{Optimal Solution (OS) Strategy.}

The solution of P3 ultimately boils down to jointly finding the best index set $\mathcal{I}^*$ of $N_\textrm{S}$ eigenvectors in (\ref{eq_eigen_modes}) and optimal power allocation $\mathbf{P}_{\mathcal{I}^*}$ to that combination of eigenvectors so that the achievable rate in (\ref{eq_new_capacity}) is maximized. 
Thus, the OS strategy for P3 is to iterate over all $\binom{N_\textrm{T}}{N_\textrm{S}}$ combinations of eigenvectors and perform power allocation for each combination whenever the channel changes. We note that for a given index set $\mathcal{I}$ and corresponding eigenvectors $\mathbf{U}_{\mathcal{I}}$, the problem P3 is concave and $\mathbf{P}_{\mathcal{I}}$  can be solved optimally using CVX \cite{cvx}. Finally, the optimal precoder is given by
\begin{align}
\label{eq_optimal_beamformer}
\mathbf{W^*} = \mathbf{U}_{\mathcal{I}^*} \sqrt{\mathbf{P}_{\mathcal{I}^*}},
\end{align}
where 
$\mathbf{U}_{\mathcal{I}^*}$ is formed from the optimal eigenvectors in (\ref{eq_eigen_modes}) and $\mathbf{P}_{\mathcal{I}^*}$ contains the optimal powers allocated to those eigenvectors. 

As the OS strategy iterates over all $\binom{N_\textrm{T}}{N_\textrm{S}}$ combinations of eigenvectors and performs power allocation for each combination, the complexity of the OS strategy is $\binom{N_\textrm{T}}{N_\textrm{S}}$ times the complexity of the power allocation subtask in P3.

\subsection{Suboptimal Solution (SS) Strategy.}

As $N_\textrm{T}$, $N_\textrm{R}$, and $N_\textrm{S}$ increase, the computational complexity of the OS strategy grows rapidly, approximately on the order of $\binom{N_\textrm{T}}{N_\textrm{S}}$, which can be further characterized using Stirling's approximation to the binomial coefficient~\cite{stirling_ref}. This may exhaust computational resources or hinder the use of the OS strategy in real-time scenarios. Instead of allocating optimal power for each $\binom{N_\textrm{T}}{N_\textrm{S}}$ eigenvector combination, we use equal power allocation for all combinations to identify $Q \le \binom{N_\textrm{T}}{N_\textrm{S}}$ candidates with the top $Q$ achievable rates in the first phase for shortlisting. In the second phase, we allocate optimal power only to these shortlisted combinations. Among these $Q$ combinations, the one yielding the highest achievable rate is selected as the suboptimal beamformer using (29). The proposed SS strategy therefore reduces the number of power allocation subtasks to $Q$ times per channel realization, significantly reducing computation cost.  The SS strategy reduces computational complexity by $(1 - Q/\binom{N_\textrm{T}}{N_\textrm{S}})\%$ relative to the OS strategy.


\section{Numerical Results}\label{section_results}

In this section, the performance of the proposed DoA-OPR-based beamforming is evaluated. Unless otherwise stated, the system parameters are stated in Table \ref{table_parameters}. In the figures, `SNR' refers to ${P}/{N_0}$. The results are obtained by averaging over multiple independent channel realizations, and the OS strategy is employed unless otherwise specified. All simulations are implemented in MATLAB, and convex optimization problems are solved using the CVX toolbox.


\begin{table}[]
\caption{Parameters for numerical evaluations.}
\label{table_parameters}
\centering
\small 
\begin{tabular}{cll}
\hline
\textbf{Parameter} & \textbf{Value} & \textbf{Description} \\ \hline
$N_\textrm{T}$ & 16 & Number of transmit antennas \\
$N_\textrm{R}$ & 8 & Number of receive antennas \\
$N_\textrm{S}$ & 4 & Number of data streams \\
$\phi$ & $45^\circ$ & True angle of departure \\
$\hat{\phi}$ & $75^\circ$ & False angle of departure \\ 
$\lambda$ & -- & Wavelength\tablefootnote{The proposed design does not depend on the absolute value of the carrier frequency.} \\
$d$ & $\lambda/2$ & Antenna spacing \\
$\kappa$ & 0 dB & Rician K-factor \\
$L$ & 20 & Number of NLoS paths \\ \hline
\end{tabular}
\end{table}

\subsection{Impact of Privacy Threshold.}
Figs. \ref{fig_privacy_pattern} and \ref{fig_privacy_rate} show the received beampattern and achievable rate, respectively,  for different privacy threshold $\gamma_\textrm{th} = \{0, 1, 2, \gamma_\textrm{max}\}$. 
In Fig. \ref{fig_privacy_pattern}, the behavior of beampatterns under different $\gamma_\textrm{th}$ values is evident. When $\gamma_\textrm{th} = 0$ (Case 3 in section \ref{sec_Problem_Formulation}), the design reduces to conventional beamforming that maximizes the achievable rate.  
For $\gamma_\textrm{th} = 1$ (Case 4 in section \ref{sec_Problem_Formulation}), equal energy is radiated toward both the true and false angles. As $\gamma_\textrm{th}$ increases beyond unity, i.e. for $\gamma_\textrm{th} = 2$ (Case 4 in section \ref{sec_Problem_Formulation}), the false angle dominates and provides location privacy to the transmitter. For $\gamma_{\textrm{th}} = \gamma_{\max}$ (Case 2 in section \ref{sec_Problem_Formulation}), the maximum power difference between the false and true directions is achieved. However, for large $\gamma_{\textrm{th}}$ values, excessive suppression of the true direction may reveal the transmitter’s location, making it undesirable. 

In Fig. \ref{fig_privacy_rate}, we can observe that the achievable rate reduces with increasing  $\gamma_\textrm{th}$ as more power is directed toward the false direction. This confirms that DoA-OPR tuning enables a trade-off between achievable rate and location privacy. Therefore, an optimum value of $\gamma_{\textrm{th}}$ can be obtained to satisfy both privacy and achievable rate requirements.

\subsection{DoA Estimation at the Receiver via Capon.}
\begin{figure}[ ]
    \centering
        \begin{subfigure}[ ]{0.48\linewidth}
        \centering
        \includegraphics[width=\linewidth]{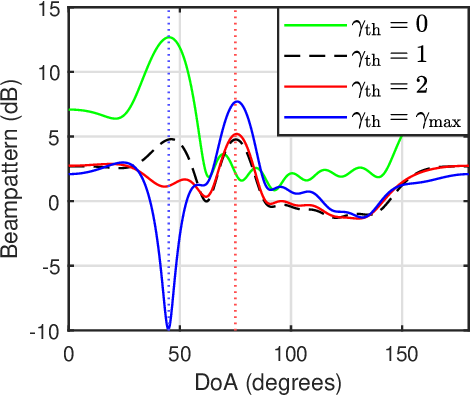}
        \caption{Beampattern at SNR = 10 dB.}
        \label{fig_privacy_pattern}
    \end{subfigure}
    \hfill    
    \begin{subfigure}[ ]{0.48\linewidth}
        \centering
        \includegraphics[width=\linewidth]{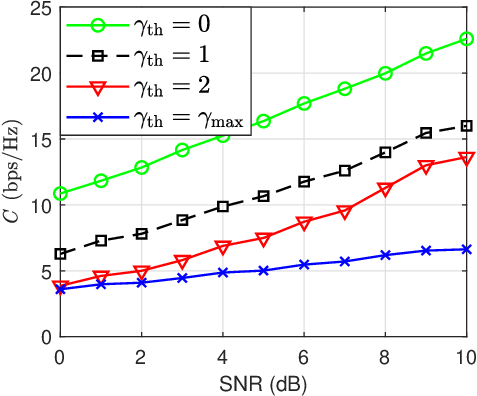}
        \caption{Achievable rate vs. SNR. 
        }
        \label{fig_privacy_rate}
    \end{subfigure}
    \caption{Impact of privacy threshold on beampattern and achievable rate. }
    \label{fig_privacy_threshold}
    \vspace{-.5cm}
\end{figure}

To verify whether the proposed DoA-OPR-based beamforming successfully achieves transmitter location privacy from the sensing-capable MIMO receiver, the receiver estimates the DoA using the Capon estimator \cite{xu2008target}. 
 The Capon estimator is applied over the beampattern shown in Fig. \ref{fig_privacy_pattern} with  64 snapshots, a diagonal loading factor of $10^{-3}$, and a spectrum resolution of \textbf{$0.5^{\circ}$}.
For $\gamma_{\text{th}} = \{0, 1, 2, \textrm{and}~  \gamma_{\max}$\},  the 
estimated DoA is $\phi_\textrm{Capon} = \{45^\circ, 45.5^\circ, 74.5^\circ, \textrm{and}~ 75^\circ$\}, respectively.
These results confirm that increasing $\gamma_{\text{th}}$ makes the false direction more dominant, validating the effectiveness of the proposed location privacy mechanism.


\subsection{Impact of Antenna Configurations on Maximal Privacy.}
\begin{figure}[ ]
    \centering
    \begin{subfigure}[t]{0.24\textwidth}
        \centering
        \includegraphics[width=\linewidth]{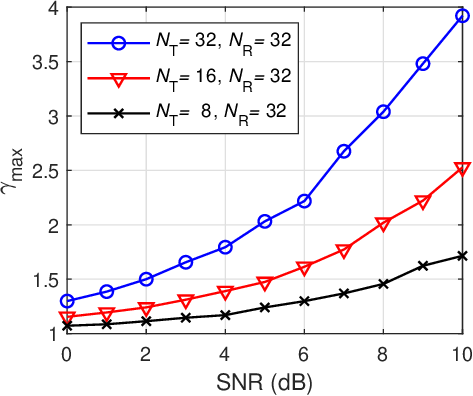}
        \caption{$N_\textrm{T}=\{8,16,32\}$, $N_\textrm{R}=32$.}
        \label{fig_antenna_Nt_daor}
    \end{subfigure}
    \hfill
    \begin{subfigure}[t]{0.24\textwidth}
        \centering
        \includegraphics[width=\linewidth]{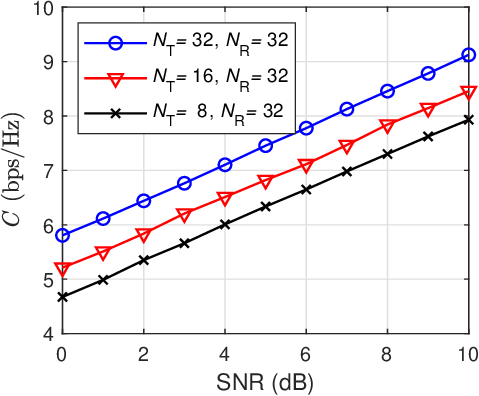}
        \caption{$N_\textrm{T}=\{8,16,32\}$, $N_\textrm{R}=32$.}
        \label{fig_antenna_Nt_rate}
    \end{subfigure}
    \vspace{0.2cm}
    \begin{subfigure}[t]{0.24\textwidth}
        \centering
        \includegraphics[width=\linewidth]{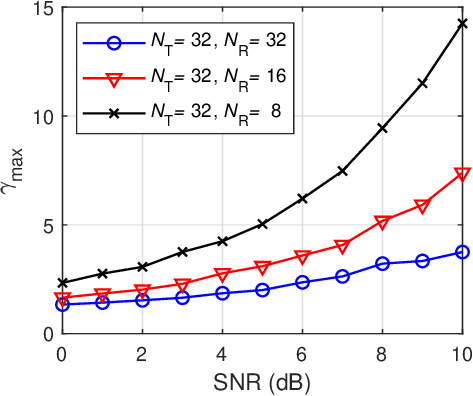}
        \caption{$N_\textrm{T}=32$, $N_\textrm{R}=\{8,16,32\}$.}
        \label{fig_antenna_Nr_daor}
    \end{subfigure}
    \hfill
    \begin{subfigure}[t]{0.24\textwidth}
        \centering
        \includegraphics[width=\linewidth]{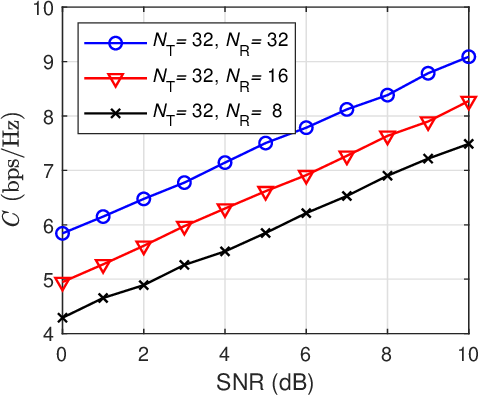}
        \caption{$N_\textrm{T}=32$, $N_\textrm{R}=\{8,16,32\}$.}
        \label{fig_antenna_Nr_rate}
    \end{subfigure}
    \caption{Impact of antenna configurations on maximal privacy and achievable rate.}
    \label{fig:antenna_config}
\end{figure}

The effect of different antenna configurations on the maximal achievable privacy $\gamma_{\max}$ is illustrated in Figs. \ref{fig_antenna_Nt_daor} and \ref{fig_antenna_Nr_daor} by varying $N_\textrm{T}$ and $N_\textrm{R}$, respectively. The corresponding achievable rate $C$ is shown in Figs. \ref{fig_antenna_Nt_rate} and \ref{fig_antenna_Nr_rate}, respectively.
As maximal privacy yields the lowest achievable rate, Figs. \ref{fig_antenna_Nt_rate} and \ref{fig_antenna_Nr_rate} show the lower bound of the achievable rate for the corresponding antenna configurations. We observe from the figures that both $\gamma_{\textrm{max}}$ and $C$ increase with SNR for a given antenna configuration. We also find that increasing $N_\textrm{T}$ and $N_\textrm{R}$ yields a better achievable rate. Comparing Figs. \ref{fig_antenna_Nt_daor} and \ref{fig_antenna_Nr_daor} we further conclude that increasing $N_\textrm{T}$ provides higher $\gamma_{\textrm{max}}$; in contrast, increasing $N_\textrm{R}$ achieves lower $\gamma_{\textrm{max}}$.







\subsection{Impact of Number of Data Streams.}
\begin{figure}[ ]
    \centering
    \begin{subfigure}[ ]{0.48\linewidth}
        \centering        \includegraphics[width=\linewidth]{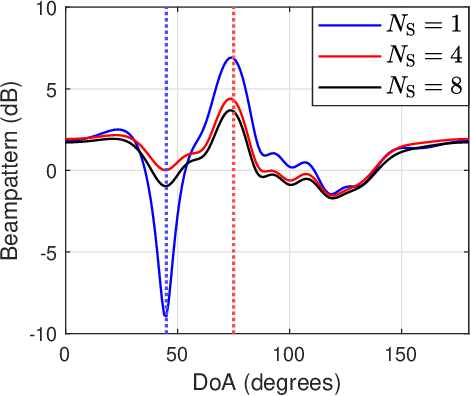}
        \caption{Beampatterns when $\gamma_\textrm{th} = 2$ and SNR = 10 dB.}
        \label{fig_streams_pattern}
    \end{subfigure}
        \hfill
    \begin{subfigure}[ ]{0.48\linewidth}
        \centering        \includegraphics[width=\linewidth]{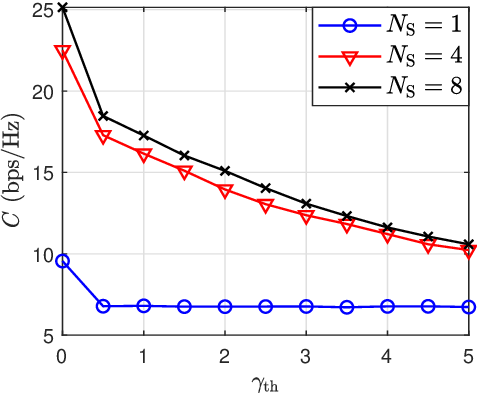}
        \caption{Achievable rate vs. $\gamma_\textrm{th}$ when SNR = 10 dB.}
        \label{fig_streams_rate}
    \end{subfigure}    
    \caption{Impact of the number of data streams on the beampattern and achievable rate.}
    \label{fig_data_streams}
\end{figure}

Figs. \ref{fig_streams_pattern} and \ref{fig_streams_rate} plot the received beampattern versus DoA and achievable rate $C$ versus privacy threshold $\gamma_\textrm{th}$, respectively, for different number of data streams $N_\textrm{S}$. 
Fig. \ref{fig_streams_pattern} shows that, as the number of data streams decreases, more power is directed towards the false direction.  As a result, this improves location privacy.
With fixed transmit power, reducing the number of data streams concentrates energy on fewer eigenmodes, allowing more power to be steered toward the false direction. This leaves less power in the true direction and leads to a lower achievable rate, as can be observed in Fig. \ref{fig_streams_rate} for a given privacy threshold.  The observation that the achievable rate decreases as the privacy threshold increases is consistent with the findings of Fig. \ref{fig_privacy_rate}.
On the other hand, more data streams enhance achievable rate; however, the incremental gains in the achievable rate diminish for a higher number of data streams.

\subsection{Performance Comparison of the OS and SS Strategies.}
\begin{figure} [] 
    \centering
    \includegraphics[width=0.5\linewidth]{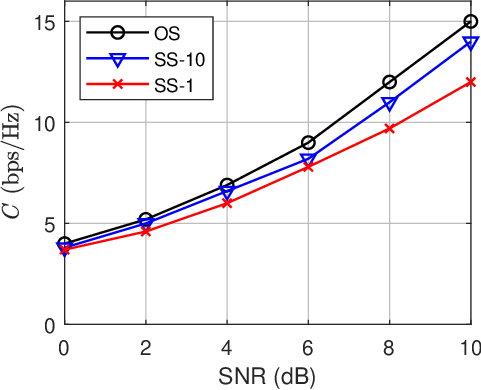}
    \caption{Performance comparison of optimal and suboptimal solution strategies when $\gamma_\textrm{th} = 2$ and $N_\textrm{S} = 4$.}
    \label{fig_eigenmode_selection}
\end{figure}

Fig. \ref{fig_eigenmode_selection} compares the achievable rate of the OS and SS strategies. In the legend, SS-10 and SS-1 represent the SS strategy with $Q = 10$ and $Q = 1$, respectively. While the OS strategy guarantees the best performance, the SS strategy achieves a near-optimal achievable rate at a moderate value of $Q$ with drastically reduced computational overhead, making it practical for real-time MIMO systems. At an SNR of 10 dB, the SS strategy with $Q = 10$ achieves nearly 85\% reduction in computation time compared to the OS strategy, with less than 7\% loss in achievable rate. This is due to the reduced number of CVX solver calls per channel realization.


\begin{figure}[ ]
    \centering
        \begin{subfigure}[ ]{0.48\linewidth}
        \centering
        \includegraphics[width=\linewidth]{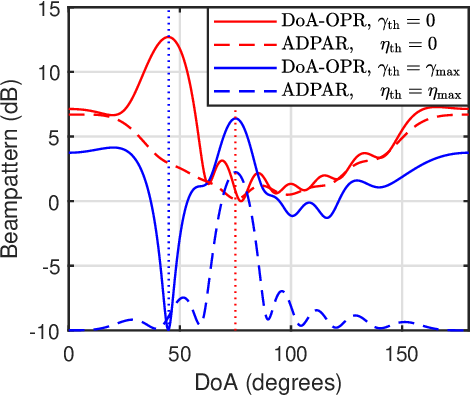}
        \caption{Beampatterns at SNR =10 dB.}
        \label{fig6a}
    \end{subfigure}
    \hfill    
    \begin{subfigure}[ ]{0.48\linewidth}
        \centering
        \includegraphics[width=\linewidth]{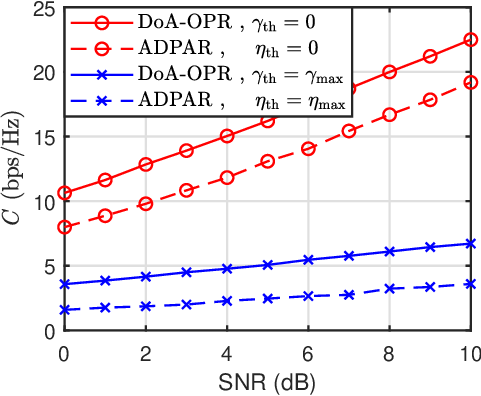}
        \caption{Achievable rate vs. SNR. 
        }
        \label{fig6b}
    \end{subfigure}
    \caption{Comparison with ADPAR-based design. }
    \label{fig6}
\end{figure}

\subsection{Performance Comparison with Existing Works.}
We compare the proposed DoA-OPR based design with the angular-domain peak-to-average ratio (ADPAR)-based design proposed in \cite{ma2024sensingResistance}. The ADPAR privacy metric achieves angular obfuscation by increasing power in the false direction relative to
the average power of the received beampattern while nulling the power in the true direction. 
As the privacy metrics are different, a fair  comparison is difficult. Therefore, in Fig. \ref{fig6} we compare the designs for two edge cases: (i)  privacy constraint is inactive ($\gamma_{\mathrm{th}} = 0$ and $\eta_{\mathrm{th}} = 0$) and (ii) achievable privacy is maximum ($\gamma_{\mathrm{th}} = \gamma_{\mathrm{max}}$ and $\eta_{\mathrm{th}} = \eta_{\mathrm{max}}$) in the respective design approaches, where $\eta_{\mathrm{th}}$ and $\eta_{\mathrm{max}}$ are the ADPAR threshold and its maximum possible value, respectively \cite{ma2024sensingResistance}.   Fig. \ref{fig6a} shows the beampattern, and the corresponding achievable rate versus SNR is shown in Fig. \ref{fig6b}. 

In both cases (i) and (ii), the proposed design achieves a higher rate with the same privacy level as validated using the Capon estimator. In case (ii), the estimated DoAs are $\phi_\text{Capon} = 74.5^{\circ} ~\textrm{and}~ 75.0^{\circ}$ for the proposed and ADPAR-based design, respectively. 
The proposed scheme achieves a higher achievable rate as the LoS component is not nulled, unlike in the ADPAR-based design.


    
\section{Conclusions}
\label{section_conclusions}
In this paper, a beamforming design is proposed to enable transmitter location privacy.
Unlike existing techniques that suppress the LoS channel component and incur considerable rate loss, the proposed design preserves the LoS path. A privacy metric termed DoA-OPR is introduced, which controls the angular domain power ratio between false and true directions. A rate-maximization problem is formulated for the beamformer design constrained on the privacy metric. The optimization problem is non-convex; however, the optimal solution is obtained through semidefinite relaxation, eigenmode selection, and corresponding optimal power allocation. 
 The Capon DoA estimator demonstrates that the proposed design achieves the same privacy level as the LoS-nulling design while providing a higher communication rate.  We also propose a computationally efficient suboptimal solution, making it suitable for real-time MIMO systems. The suboptimal design achieves nearly 85\% reduction in computation time with less than 7\% loss in achievable rate as compared to the optimal strategy at an SNR of 10 dB.


\section*{Acknowledgment}
This work was supported in part by Taighde Éireann – Research Ireland under Grant numbers 22/PATH-S/10788 and 24/FFP-P/12895.
\vspace{-.2cm}
\bibliographystyle{IEEEtran}
\bibliography{references}

\end{document}